\documentclass[review,10pt,2022, 3p]{elsarticle}

\usepackage[utf8]{inputenc}
\usepackage[T1]{fontenc}
\usepackage{lscape}
\usepackage{caption}
\usepackage{subcaption}
\usepackage{booktabs}
\usepackage{comment}
\usepackage{wrapfig}
\usepackage{graphicx}
\usepackage[export]{adjustbox}
\usepackage{amsmath}
\usepackage{float}
\usepackage[super]{nth}
\usepackage[dvipdfmx]{media9}
\usepackage{enumitem}

\usepackage{xcolor}

\usepackage{todonotes}      
\usepackage[pagewise]{lineno}

\usepackage{hyperref}
\hypersetup{
    colorlinks=true,
    linkcolor=blue,
    filecolor=magenta,      
    urlcolor=blue
}

\usepackage{orcidlink}

\usepackage{doi}

\usepackage[acronym]{glossaries}

\makeglossaries

\newacronym{osm}{OSM}{OpenStreetMap}
\newacronym{cc}{CC}{Carrying Capacity}
\newacronym{pcc}{PCC}{Physical Carrying Capacity}
\newacronym{rcc}{RCC}{Real Carrying Capacity}
\newacronym{ecc}{ECC}{Effective Carrying Capacity}
\newacronym{tcc}{TCC}{Tourism Carrying Capacity}
\newacronym{api}{API}{Application Programming Interface}
\newacronym{xml}{XML}{Extensible Markup Language}
\newacronym{abm}{ABM}{Agent-Based Modeling}
\newacronym{sfm}{SFM}{Social Force Model}
\newacronym{utm}{UTM}{Universal Transpose Mercator}
\newacronym{los}{LOS}{Level Of Service}
\newacronym{gis}{GIS}{Geographic Information System}
\newacronym{vm}{VM}{Virtual Machine}
\newacronym{fccn}{FCCN}{Fundação para a Computação Científica Nacional}
\newacronym{incd}{INCD}{Infraestrutura Nacional de Computação Distribuída}
\newacronym{unwto}{UNWTO}{World Tourism Organization}

\journal{arXiv}

\begin{document}
\begin{frontmatter}


\title{A Carrying Capacity Calculator for Pedestrians Using OpenStreetMap Data:\\
        Application to Urban Tourism and Public Spaces}

\author[1]{Duarte Sampaio de Almeida\orcidlink{0000-0001-5459-4113}}
\ead{dsbaa@iscte-iul.pt}

\author[1]{Rodrigo Simões\orcidlink{0009-0007-9303-8954}}
\ead{rjbss@iscte-iul.pt}

\author[1]{Fernando Brito e Abreu\orcidlink{0000-0002-9086-4122}}
\ead{fba@iscte-iul.pt}

\author[1]{Adriano Lopes\orcidlink{0000-0003-3685-0659}}
\ead{adriano.lopes@iscte-iul.pt}

\author[2,3]{Inês Boavida-Portugal\orcidlink{0000-0001-9932-9241}}
\ead{iboavida-portugal@campus.ul.pt}

\affiliation[1]{
    organization={ISTAR-IUL, Instituto Universitário de Lisboa (Iscte-IUL)},
    city={Lisbon},
    country={Portugal}
}

\affiliation[2]{
  organization={University of Lisbon, Centre of Geographical Studies, Institute of Geography and Spatial Planning},
  city={Lisbon},
  country={Portugal}
}

\affiliation[3]{
  organization={Associated Laboratory Terra},
  country={Portugal}
}

\begin{abstract}
Determining the carrying capacity of urban tourism destinations and public spaces is essential for sustainable management. This paper presents an online tool that calculates pedestrian carrying capacities for user-defined areas based on OpenStreetMap (OSM) data. The tool considers physical, real, and effective carrying capacities by incorporating parameters such as area per pedestrian, rotation factor, corrective factors, and management capacity. The carrying capacity calculator aids in balancing environmental, economic, social, and experiential factors to prevent overcrowding and preserve the quality of life for residents and visitors. This tool is particularly useful for tourism destination management, urban planning, and event management, ensuring positive visitor experiences and sustainable infrastructure development. We detail the implementation of the calculator, its underlying algorithm, and its application to the Santa Maria Maior parish in Lisbon, highlighting its effectiveness in managing urban tourism and public spaces.
\end{abstract}


\end{frontmatter}

\section{Introduction}

Determining the carrying capacity is essential for the sustainable management of tourism destinations and public urban spaces, balancing environmental, economic, social, and experiential factors. Knowing the carrying capacity helps minimize environmental impact by limiting visitor numbers to levels that do not degrade natural resources and ecosystems, thereby preserving biodiversity and heritage sites. Economically, it ensures long-term viability by preventing resource overuse, which can lead to economic downturns if destinations become less attractive due to degradation. Socially, it maintains the quality of life for residents by avoiding overcrowding and overuse of public services and infrastructure, while protecting local cultures from being overwhelmed by excessive tourism or events.

Visitor experience is enhanced through carrying capacity management, ensuring positive experiences, free from long wait times, traffic congestion, and insufficient facilities, and maintaining safety standards to prevent accidents and hazards. Additionally, it guides sustainable infrastructure development and maintenance, preventing overburdening and managing costs associated with repairing overused facilities.
Carrying capacity informs regulatory and planning efforts, aiding policymakers in creating effective regulations and guidelines, and assisting urban planners and event managers in designing spaces that sustainably accommodate intended visitor numbers. It provides a feedback mechanism for continuous monitoring and adaptive management strategies to address emerging issues, enabling data-driven decisions that enhance sustainability.

A carrying capacity calculator for pedestrians is, therefore, an important tool in several fields such as tourism destination management, urban management, events management (e.g. concerts, music festivals, political rallies, fireworks, plays), and attractions management (e.g. monuments, museums, theme parks, archaeological sites).

The online tool described in this paper calculates the carrying capacities for pedestrians in a user-defined area defined interactively upon \gls{osm}, an open data geographic database with worldwide coverage, updated and maintained collaboratively by a large community of volunteers following a peer production model such as the one of Wikipedia \cite{Haklay2008}.

This paper is organized as follows: in the next section we provide some background on the \gls{osm} initiative and on carrying capacity concepts; then, in section \ref{sec:reqspec} we define the functional and non-functional requirements of a carrying capacity calculator; in section \ref{sec:algorithm} we describe the algorithm implemented to determine the carrying capacity upon an \gls{osm} map; in section \ref{sec:ccc} we present the user interaction steps with tool and, finally, on section \ref{sec:conclusion} we provide some final remarks. 

\section{Background}

\subsection{The OpenStreetMap initiative} 
\label{subsec:OSM}

The \gls{osm} is a worldwide open data geographic mapping project developed, since 2004, through the contribution of a large community of volunteers spread across the globe. The mapping elements are stored in a geographic database which can be used without limitations, as long as credit is given. Currently \gls{osm} has a wide coverage of the earth's surface and major cities are represented with significant accuracy. The data gathered, stored, and served by the project consists of an immense set of geographic entities, referenced in \gls{osm}'s data model as elements of the following three main types:

\begin{itemize}[topsep=0pt, partopsep=0pt, parsep=0pt, itemsep=0pt]
 
    \item \textbf{Node}: Represents a specific point defined by its coordinates (latitude and longitude) and an identifier. Nodes can define standalone point features, such as points of interest, and street elements (e.g. benches, trees, statues), or define points on a way or relation.
    
    \item \textbf{Way}: A way represents a multi-point geometry, consisting of an ordered list of nodes, with a minimum of 2 nodes, and a maximum of 2,000 nodes. Ways allow defining linear or multi-linear elements (e.g., roads, streets, water streams, public transport paths), and areas (e.g., buildings, squares, gardens, parks).
    
    \item \textbf{Relation}: Represents an ordered list of nodes, ways, or relations, defining a relationship between its members. The geographic entities modeled by a relation include multi-polygons (a group of polygons considered as a single entity), administrative boundaries, routes, etc.
 
\end{itemize}

\begin{figure}[!ht]
\centering
\includegraphics[width=0.75\textwidth]{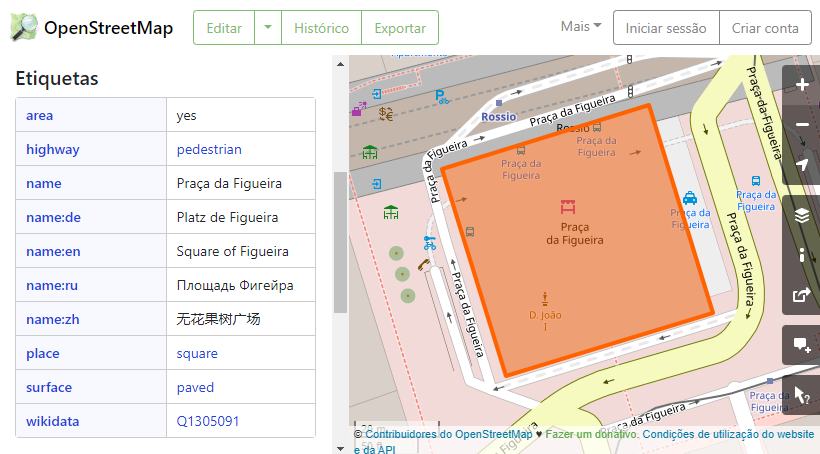}
\caption{An \gls{osm} page with its tag-value mapping}
\label{fig:osm}
\end{figure}

Elements are characterized by tags, i.e. key-value pairs that describe specific features. The key describes a topic, category, or feature type, while the value provides detail for the key-specified feature.
The tagging system follows an agreed convention to allow for accurate characterization of elements. In figure \ref{fig:osm}, the element representing the central pavement of the Praça da Figueira square (Lisbon, Portugal) is selected in the map of the \href{https://www.openstreetmap.org/}{\gls{osm} website}, showing its tags to the left.
 
\gls{osm} data can be accessed mainly via the following sources:

\begin{itemize}[topsep=0pt, partopsep=0pt, parsep=0pt, itemsep=0pt]

    \item \href{https://planet.openstreetmap.org/}{Planet.osm file}: A file containing all existing information on the \gls{osm} database. It can be downloaded via torrents or mirrors.
    
    \item \href{https://www.openstreetmap.org/export}{Export files}: Files obtained through the \gls{osm}'s main website, which allows selecting all elements within a latitude-longitude bounding box.
    
    \item \href{https://dev.overpass-api.de/overpass-doc/en/}{Overpass API}: An Application Programming Interface (API) that allows for custom queries to obtain specific types of elements.
    
    \item \textbf{Third-party providers}: Entities, external to the mapping project, which provide raw, derived, or pre-processed \gls{osm} data.
    
\end{itemize}

Files can be formatted in plain XML, or PBF (Protocolbuffer Binary Format), a file format designed to be more efficient and faster to parse. PBF files contain the same geographic data as XML files but occupy less space. 

\subsection{Carrying capacity} 
\label{subsec:CarryingCapacity}

The concept of \gls{cc}, originally from the field of ecology, is defined as the maximum supported population number for a given environment, until there are no sufficient resources to maintain population growth, and degradation starts to occur. One of the first usages of this concept in the Malthusian population theory \cite{Seidl1999}, which states the exponential growth of the human population is limited to existing resources (which cannot grow exponentially), turning the growth model into a logistic model, being its upper limit the \gls{cc} of the human population.
The concept has been applied in multiple other fields, such as economy, biology, and population studies \cite{Kennell2014}, and has had various definition iterations according to its application.

The concept has been approached in tourism studies, being known as \gls{tcc}.  
According to the \href{https://www.unwto.org/}{UN World Tourism Organization (UNWTO)}, it is defined as \textit{“the maximum number of people that may visit a tourist destination at the same time, without destroying the physical, economic, social-cultural environment and an unacceptable decrease in the quality of visitors’ satisfaction”} \cite{WTO1981}. Getz \cite{Getz1983} stated there are 6 \gls{tcc} categories: physical, economic, perceptive, social, ecological, and political. Respectively, each category represents the maximum usage of a tourist resource before physical deterioration, excessive economic dependence, perception of overcrowding by tourists, ecological damage to nature, and political instability.

There is not a consensus on the usefulness of numerical values for \gls{tcc}. McCool et al. \cite{McCool2001} assert that these values are "magic values", dependent on many realistically hard-to-model variables, which can lack stability, and are inappropriate for solving the problems of tourism development. Nonetheless, these authors acknowledge the usefulness of previous \gls{tcc} research for managing tourist sites. In other words, while the \gls{tcc} is considered an imperfect measure for the actual \gls{cc} of a touristic place of interest, it can be used as a reference value to measure overcrowding and compare scenarios.

We now describe the three levels of \gls{tcc} of the seminal methodology proposed in \cite{Cifuentes1992}.

\subsubsection{\gls{pcc}}

This level of \gls{tcc} represents the maximum number of visitors that can be physically accommodated into an area over some time. It is calculated using the following formula:
    \[PCC = \frac{A}{Ap} \times Rf\]
    where \textit{A} is the available area for tourism activity, \textit{Ap} is the area used per tourist for a specific type of tourism activity, and \textit{Rf} is the rotation factor, which is the number of permissible visits over a specified time (Usually calculated by daily open hours) and calculated by dividing the amount of time usable in the day for tourists by the meantime of a visit.

Examples of calculating the \emph{rotation factor} can be found in \cite{Insani2020, Jangra2021, Putri2021, Bera2023}.

\subsubsection{\gls{rcc}}

This value is derived from a previously calculated \gls{pcc}, applying corrective factors with different natures (e.g., physical, ecological, economical), specific to the location.  The formula to assess \gls{rcc} is:
        \[RCC = PCC \times (cf1 \times cf2 \times … cfn)\]
    where \textit{cf1...cfn} are the defined corrective factors. Each factor is calculated using the expression:        
        \[cf = 1 - \frac{Lm}{Tm}\]
    where \textit{Lm} is the limiting magnitude of the factor, and \textit{Tm} is the total magnitude of the factor.

Illustrative examples of the application of corrective factors (aka as ``correction factors'') can be found in \cite{Suwarno2018, deVera2019, Jangra2021, Putri2021, Petronijevic2022, Bera2023}.

\subsubsection{\gls{ecc}}

The final level of \gls{tcc} is derived from the \gls{rcc}, multiplying by  
        \[ECC = RCC \times Mc\]
    where \textit{Mc} is the management capacity of the site. This variable is often determined by the adequacy of the available infrastructure, equipment, and staff for the tourism activity.
    
Regarding management capacity (a percentage), the seminal work on \gls{tcc} \cite{Cifuentes1992} advocates we should consider numerous variables such as the juridical circumstances, policies, protocols, apparatus, peoples, capital, infrastructure, and facilities. Estimation examples can be found in \cite{Insani2020, Putri2021, Bera2023}.

As for the ideal area/space per tourist/pedestrian, this complex cultural and activity-dependent issue will be detailed in the following subsection.

\subsection{Tourism proxemics} 
\label{subsec:Proxemics}

Several authors have observed that different cultures perceive crowding differently \cite{Evans2000, Kaya2003}. In addition, each tourist activity may require a different area. For example, the area required by a tourist sunbathing on a beach towel is greater than that required by the same tourist in a restaurant, walking downtown, or watching fireworks. The estimation of this variable has been addressed by several authors such as in \cite{Itami2002, Sahani2017, deVera2019, Meligi2021} in the framework of a research area known as ``proxemics''. \cite{McCall2017}. The latter refers to the study of how people use and perceive personal space, a concept introduced in \cite{Hall1968}. The latter focuses on the physical distances maintained between a person and others, which can vary based on cultural norms, the type of interaction, and the specific setting, such as crowded attractions versus open spaces.

Cultural differences play an important role, as norms of personal space vary across cultures, affecting interactions and experiences at destinations. Tourist proxemics should also examine social interactions, such as how tourists interact with locals, other tourists, and service providers, influencing behavior in queues, at attractions, during tours, and in accommodations. The impact of spatial behavior on the overall tourist experience is another crucial aspect, as factors such as crowding can significantly affect enjoyment and perception.

Finally, the design and management of tourist spaces are informed by understanding proxemics, helping to enhance comfort, flow, and satisfaction. Overall, studying tourist proxemics aids in creating culturally sensitive and enjoyable visitor experiences by effectively managing personal space dynamics.

\section{Requirements specification}
\label{sec:reqspec}

This section describes the requirements of the carrying capacity calculator using the user stories format \cite{Beck2000}. The latter are short, simple descriptions of a feature told from the perspective of the person who desires the new capability, capturing requirements concisely and understandably, focusing on the value delivered to the user rather than detailed technical specifications, thereby ensuring that development efforts are aligned with user needs and priorities.

\subsection{Functional user stories}

As a user, I want to:

\begin{itemize}[topsep=0pt, partopsep=0pt, parsep=0pt, itemsep=0pt]

    \item zoom in and out on a world map based on OpenStreetMap;
  
    \item select interactively an area in the map or by providing the corresponding GeoJSON;

    \item observe the usable area for pedestrians in the currently selected area, and measure it in square meters and its percentage of the total selected area; the usable area for pedestrians is usually a collection of disconnected polygons, herein called "pedestrian spaces";
 
    \item obtain the Physical Carrying Capacity (PCC), Real Carrying Capacity (RCC), and Effective Carrying Capacity (ECC), as proposed in \cite{Cifuentes1992}, for the selected area, by providing the required input parameters (area per pedestrian, rotation factor, corrective factors, and management capacity) with adequate default values;

    \item see the border of the pedestrian spaces
    
    \item see contextual help whenever I am required to provide input values

\end{itemize}

\subsection{Non-functional user stories}

As a user, I want to:

\begin{itemize}[topsep=0pt, partopsep=0pt, parsep=0pt, itemsep=0pt]

    \item be aware of the progress of my requests through a progress indicator;

    \item obtain the results of my requests in a reasonable time;

\end{itemize}

\section{Algorithm for the available area for tourism activity}
\label{sec:algorithm}

\begin{figure}[htp]
     \centering
     \begin{subfigure}[t]{0.47\linewidth}
         \centering
         \includegraphics[width=\linewidth, frame]{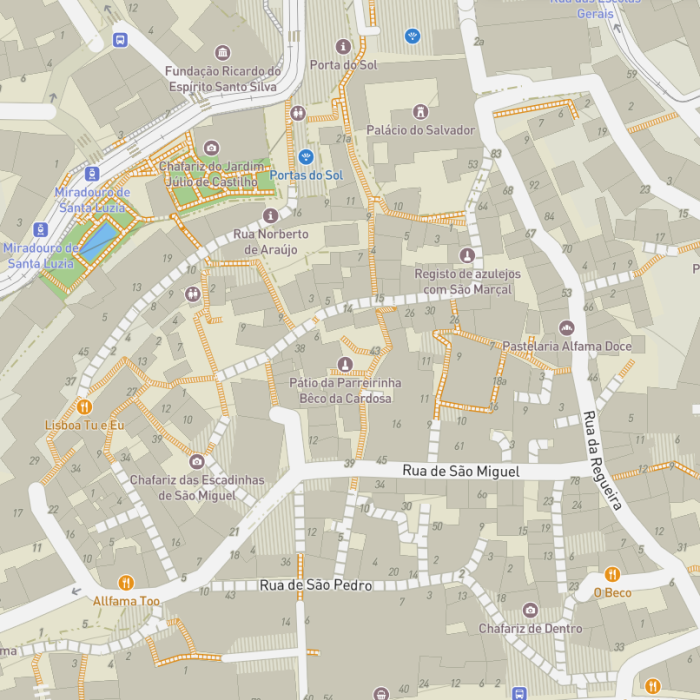}
         \caption{The XII century Moorish Alfama neighborhood}
         \label{fig:alfama}
     \end{subfigure}
     \qquad
     \begin{subfigure}[t]{0.47\linewidth}
         \centering
         \includegraphics[width=\linewidth, frame]{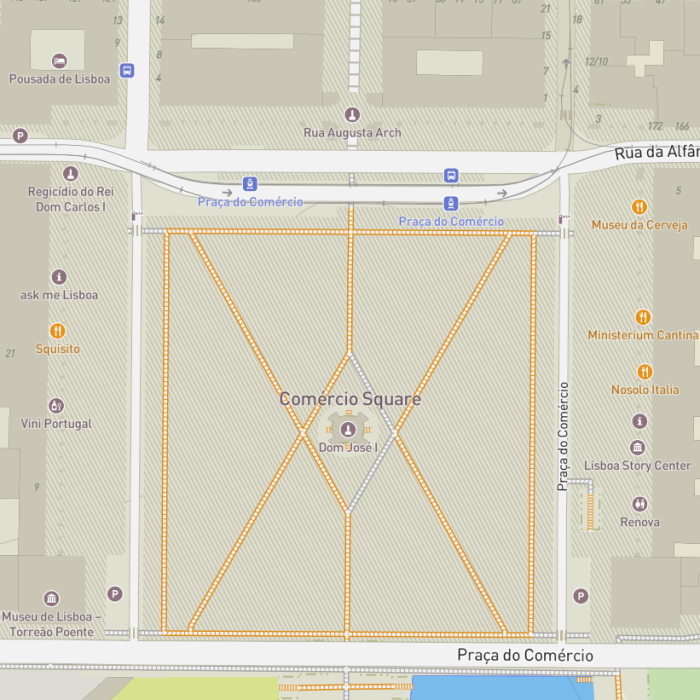}
         \caption{The XVIII century Terreiro do Paço square}
         \label{fig:terreiro_paco}
     \end{subfigure}
        \caption[b]{Influence of urban fabric on carrying capacity (two examples from Lisbon, Portugal)}
        \label{fig:urban_fabric}
\end{figure}

The notion of ``available area'' for tourism activity is central to calculating \gls{tcc}. In city scenarios, that area depends on the urban fabric, as shown in the two Lisbon tourism attraction areas in figure \ref{fig:urban_fabric}.

The algorithm takes as input a bounding line (e.g. a polygon), a list of \gls{osm} features within the boundary, all in GeoJSON format, and an options object, and computes the corresponding available area. It then removes the georeferenced features that are not considered accessible to pedestrians from the bounded area using the \textit{difference} function from \textit{Turf.js}, leaving only the accessible area. If the features are one- or two-dimensional (e.g. points, line strings), a buffer of the size specific to the feature type is added to convert them to polygons.

The main "unavailable" features are buildings, water, roads, railways, restricted areas (e.g. airways, military zones), urban furniture, trees, small monuments (e.g. monuments that are not polygonal buildings), and barriers. More types will be added over time, as \gls{osm} has many tags and different feature types.

Its algorithm implementation uses the \href{https://www.w3schools.com/html/html5_webworkers.asp}{Web Workers API} for parallelization, making the computations much faster than in single-threaded mode.

\section{The carrying capacity calculator tool}
\label{sec:ccc}

\begin{figure}[H]
    \centering
    \includegraphics[width=0.9\linewidth, frame]{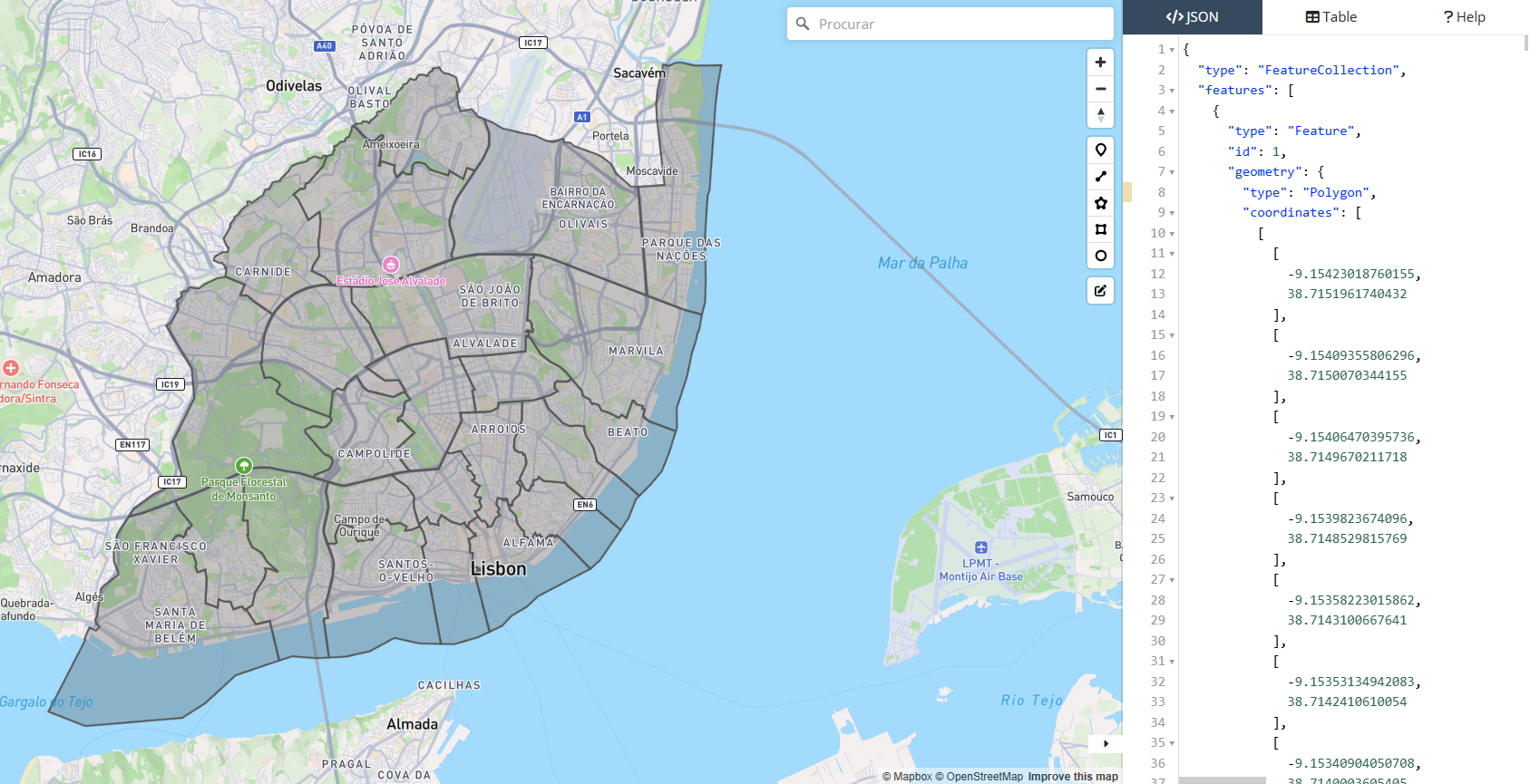}
    \caption{All Lisbon parishes: the GeoJSON code provided by the \href{https://lisboaaberta.cm-lisboa.pt/index.php/pt/informacao-de-base-e-cartografia}{Lisboa Aberta} initiative was pasted in the right panel}
    \label{fig:lisbon_parishes}
\end{figure}

\begin{figure}[H]
    \centering
    \includegraphics[width=0.9\linewidth, frame]{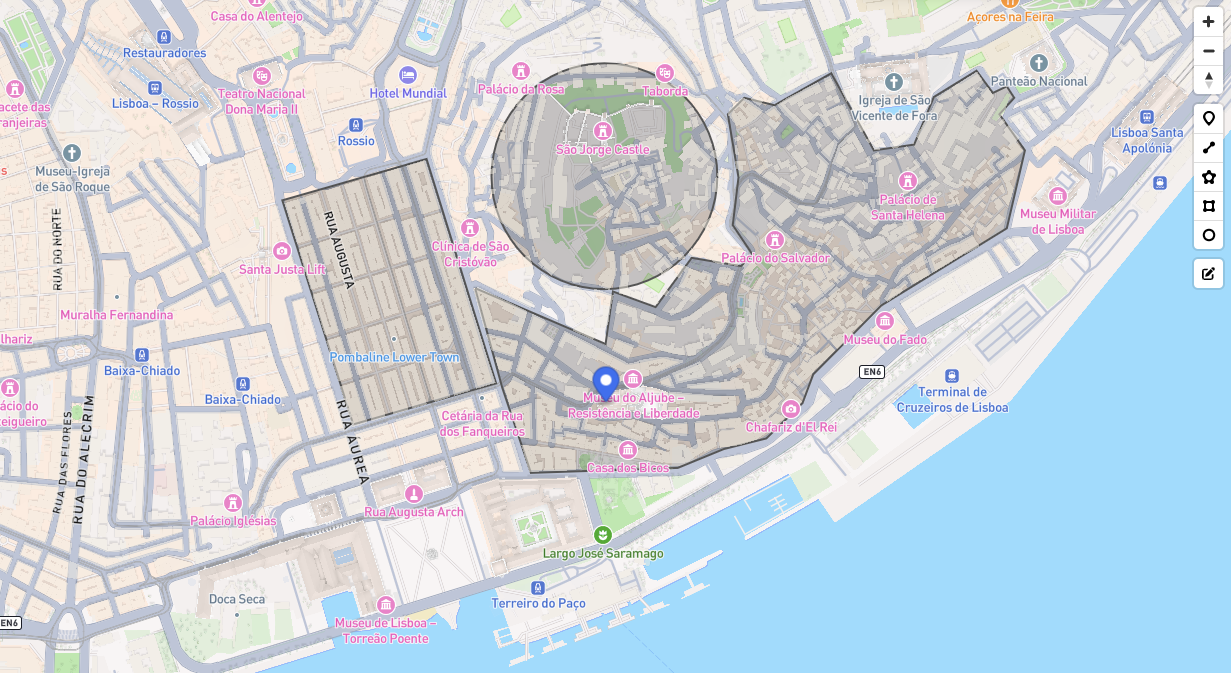}
    \caption{User-selected areas in Santa Maria Maior parish, using the shape tools on the vertical toolbar}
    \label{fig:selected_areas}
\end{figure}

This tool is an extended version of the \textit{geojson.io} web app, where the user can parameterize the previously described algorithm to obtain the three levels of \gls{tcc}. The target areas may be selected as follows:

\begin{itemize}[topsep=0pt, partopsep=0pt, parsep=0pt, itemsep=0pt]

    \item user pastes GeoJSON code in the corresponding panel on the right, as shown in figure \ref{fig:lisbon_parishes};
    
    \item user uses one of the available shape tools (circle, rectangle, and polygon) as shown in figure \ref{fig:selected_areas}.
    
\end{itemize}

\begin{figure}
    \centering
    \includegraphics[width=\linewidth, frame]{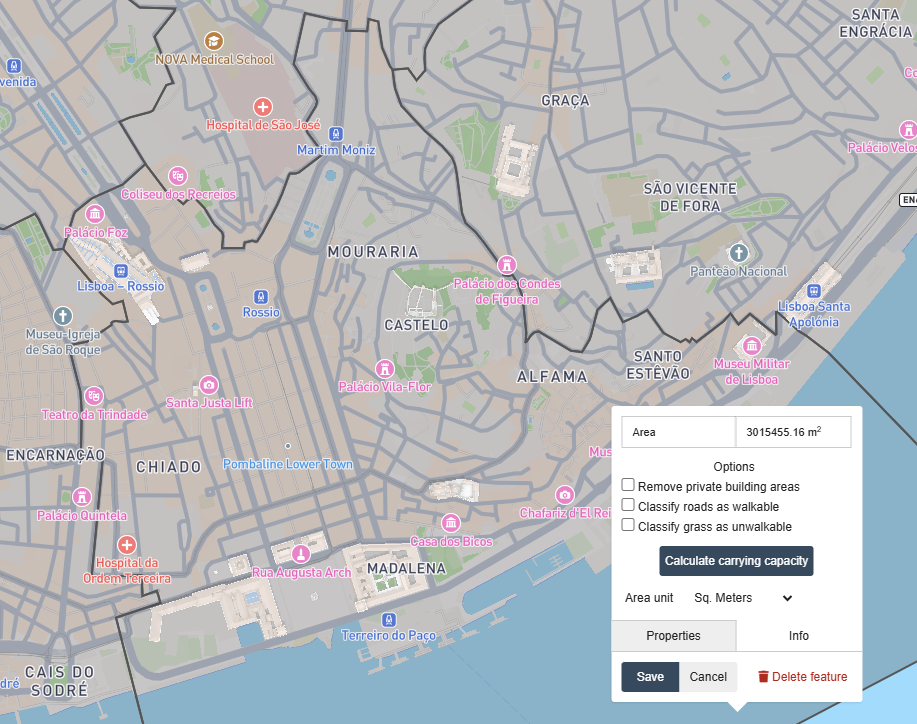}
    \caption{Available options before triggering \gls{tcc} calculation}
    \label{fig:options}
\end{figure}

After selecting an area, an options panel allows a more flexible calculation of the available area. The currently available options, shown in figure \ref{fig:options} are the following:

\begin{itemize}[topsep=0pt, partopsep=0pt, parsep=0pt, itemsep=0pt]

    \item Remove inner areas within buildings that may not be publicly accessed;
    
    \item Classify roads as walkable, to account for road blockages during special events like a marathon;
    
    \item Classify grass as not walkable, to remove vegetation spaces possibly restricted for pedestrian stepping.
    
\end{itemize}

\begin{figure}
    \centering
    \includegraphics[width=0.8\linewidth, frame]{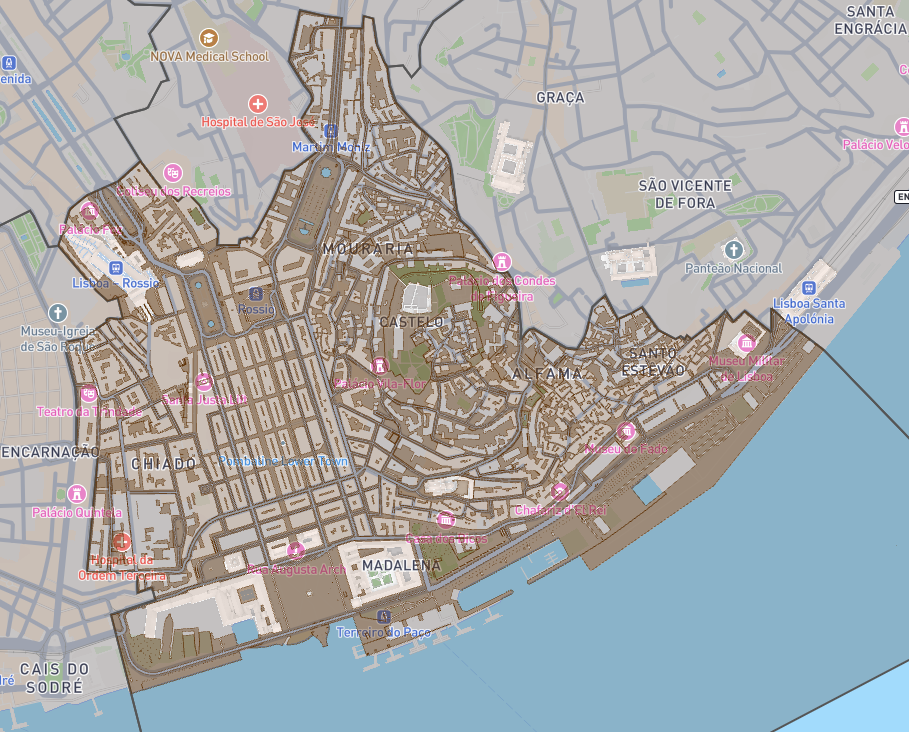}
    \caption{Walkable areas in Santa Maria Maior parish shown in brown}
    \label{fig:walkable_area}
\end{figure}

\begin{figure}
    \centering
    \includegraphics[width=\linewidth, frame]{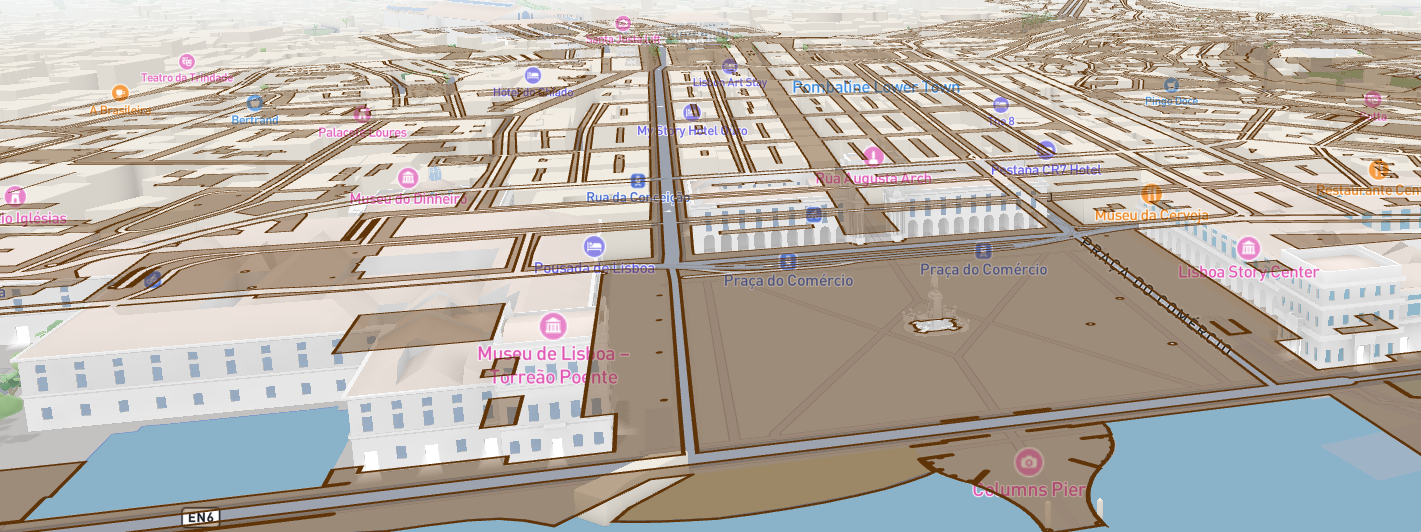}
    \caption{Perspective view of Lisbon downtown with walkable areas shown in brown}
    \label{fig:perspective}
\end{figure}

When the user presses the black-highlighted \emph{``Calculate carrying capacity''} button in figure \ref{fig:options}, the tool uses the Overpass API to retrieve the features within the selected area (by clicking on it), converts them to GeoJSON format, and passes them to the algorithm described in section \ref{sec:algorithm} that calculates the walkable area. Since the corresponding computation can take a considerable time for large areas, a percentage of completion indicator is shown for the user to monitor progress, a requirement specified in section \ref{sec:reqspec}. When this more intensive processing is done, the walkable areas are rendered in brown, as in figures \ref{fig:walkable_area}, and \ref{fig:perspective}.

\begin{wrapfigure}{r}{0.3\textwidth}
     \centering
     \includegraphics[width=\linewidth, frame]{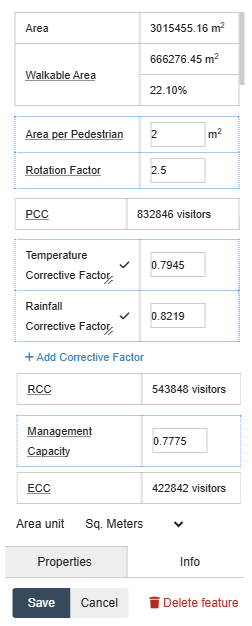}
     \caption{\gls{tcc} parametrization and results panel}
     \label{fig:parametrization}
\end{wrapfigure}

When the walkable area is calculated, a popup panel like the one displayed in figure \ref{fig:parametrization} displays the statistics the values of the total walkable area and its overall percentage), and input fields where the user can enter the area per pedestrian (see section \ref{subsec:Proxemics}), the rotation factor, custom correction factors, and the management capacity (see section \ref{subsec:CarryingCapacity}), and obtain the values of the various stages of \gls{tcc}. We will now describe how the parameters were determined.

\subsection{Area per Pedestrian}
We should keep in mind that in tourist areas it is often preferable to maintain lower densities to enhance the visitor experience, allowing for easier movement, sightseeing, and overall enjoyment. The quality of space, including seating, amenities, and pedestrian flow, also plays a significant role in how density is perceived. Therefore, we considered 2 square meters per pedestrian to foster a comfortable experience where tourists can move freely, take pictures, and enjoy the space without feeling rushed or cramped. 

\subsection{Rotation Factor}
The time available to tourists in a day to visit the parish of Santa Maria Maior is about 10 hours (number of commercial opening hours - from 9 a.m. to 7 p.m.) and the average duration of a visit is 4 hours. Therefore, \emph{Rf = 10/4 = 2.5}.

\subsection{Corrective Factors}
Although the tool allows for an arbitrary number of correction factors, for the sake of illustration we have considered only two, one related to temperature and another to precipitation. Based on data from the \href{https://www.ipma.pt/}{Portuguese Institute for Sea and Atmosphere (IPMA)}, we considered that Lisbon experiences, on average, about 290 days per year where the minimum temperature does not fall below 10 degrees Celsius, so the temperature correction factor is 290/365 = 79.45\%.
Also based on IPMA data, we considered that the average number of days when the rainfall exceeds 3mm is 65, so the rainfall correction factor is (365-65)/365 = 82.19\%.
    
\subsubsection{Management Capacity}
Estimating this parameter (a percentage) for the parish of Santa Maria Maior in Lisbon would require a comprehensive analysis of several factors, as described in the seminal work on \gls{tcc} \cite{Cifuentes1992}. Since we are concerned here with presenting a plausible value, we decided to use foundational Large Language Models (LLMs). To this end, we presented the following prompt to \emph{OpenAI's ChatGPT} and \emph{Microsoft's Copilot}:

\vspace{\baselineskip}  
\begin{minipage}{\linewidth}
  \colorbox{gray!10}{\parbox{0.9\linewidth}{
    \texttt{The management capacity of a tourist destination (a percentage) expresses the adequacy of the available infrastructure, equipment, and personnel for the tourist activity. The seminal work on Tourism Carrying Capacity [Cifuentes1992] advocates that we should consider numerous variables such as juridical circumstances, policies, protocols, apparatus, people, capital, infrastructure, and facilities. Please estimate the management capacity of the parish of Santa Maria Maior in Lisbon.}
  }}
\end{minipage}
\vspace{\baselineskip}

After some prompt engineering, ChatGPT answered 75.5\% and Copilot answered 80\%, so we took the average, i.e. 77.75\%.

\section{Conclusion}
\label{sec:conclusion}

We developed an online tool to visualize and calculate available (or walkable) outdoor space for pedestrian use and the three stages of \gls{tcc}. The application is an extension of the \textit{geojson.io} tool, allowing a user to select GeoJSON areas on a map and calculate the walkable area within those geometries using OpenStreetMap open georeferenced data and the Turf.js geoprocessing library. The resulting walkable area is then used to calculate carrying capacity levels, given user input. 
The core calculations of the application are performed locally, requiring only \gls{osm} data fetches. The Web Workers API allows multiple CPU cores to be used in parallel for faster calculations, making the web application highly scalable.

This capacity calculator, the source code for which is available on \href{https://github.com/resetting-eu/Carrying_Capacity_Calculator/}{GitHub}, can be invaluable to a variety of professions and roles due to its wide range of applications. Event planners and organizers can use it to plan and manage the layout of venues to ensure safe and comfortable occupancy levels for concerts, festivals, and gatherings. Emergency response coordinators can use it to effectively plan and manage evacuation procedures and shelter capacity. Architects and safety inspectors can use the calculator to design spaces and verify compliance with safety standards and occupancy limits.

Public transportation planners will find it useful for managing and planning the capacity of stations, platforms, and vehicles to ensure smooth and safe operations. Facility managers can manage the capacity of buildings, such as shopping malls, and maintain safety during busy periods. Public space designers can create parks, plazas, and other public areas that accommodate visitors comfortably and safely.

Recreation managers can ensure that parks, beaches, and other recreational areas do not exceed safe occupancy levels. Sports venue managers can manage seating and standing areas in stadiums and arenas to improve safety and the spectator experience. Theme park designers and operators can use the calculator to ensure that attractions and queuing areas can safely handle peak crowds. Exhibition organizers can design spaces that maximize the visitor experience while maintaining safety standards.

Finally, city event coordinators can plan citywide events, parades, marathons, and public demonstrations with proper crowd management to ensure the safety and enjoyment of all participants. This calculator provides a critical tool for these professionals to improve safety, efficiency, and overall event quality.

We invite the reader to explore this tool, available online at \href{http://carryingcapacity.eu/}{http://carryingcapacity.eu/}.

\section*{Acknowledgments}

This work has been developed by the \href{https://sites.google.com/iscte-iul.pt/resetting-project}{Iscte-IUL team} participating in the \href{https://www.resetting.eu/}{RESETTING project}, funded by the \href{https://single-market-economy.ec.europa.eu/smes/cosme_en}{European COSME program}, managed by \href{https://eismea.ec.europa.eu/index_en}{EISMEA (European Innovation Council and SMEs Executive Agency)}, under the COS-TOURINN 101038190 grant. It was also partially funded by the Portuguese FCT, under projects UIDB/04466/2020 and UIDP/04466/2020. INCD, the Portuguese National Distributed Computing Infrastructure, provided the cloud infrastructure used, funded by FCT and FEDER under project 01/SAICT/2016 \#022153.

\bibliographystyle{ACM-Reference-Format}

\bibliography{references} 

\end{document}